\documentclass[letterpaper]{article} % DO NOT CHANGE THIS
\usepackage{my2026}  % DO NOT CHANGE THIS
\usepackage{times}  % DO NOT CHANGE THIS
\usepackage{helvet}  % DO NOT CHANGE THIS
\usepackage{courier}  % DO NOT CHANGE THIS
\usepackage[hyphens]{url}  % DO NOT CHANGE THIS
\usepackage{graphicx} % DO NOT CHANGE THIS
\usepackage{natbib}  % DO NOT CHANGE THIS AND DO NOT ADD ANY OPTIONS TO IT
\usepackage{caption} % DO NOT CHANGE THIS AND DO NOT ADD ANY OPTIONS TO IT
\usepackage{algorithm}
\usepackage{algorithmic}
\usepackage{arydshln}
\usepackage{amsmath}
\usepackage{amsfonts}
\usepackage{subcaption}
\usepackage{bm}
\usepackage{booktabs}
\usepackage{multirow}
\usepackage{microtype}

\usepackage{newfloat}
\usepackage{listings}
\DeclareCaptionStyle{ruled}{labelfont=normalfont,labelsep=colon,strut=off} % DO NOT CHANGE THIS
\floatstyle{ruled}
\newfloat{listing}{tb}{lst}{}
\floatname{listing}{Listing}
\title{\includegraphics[height=2.4em]{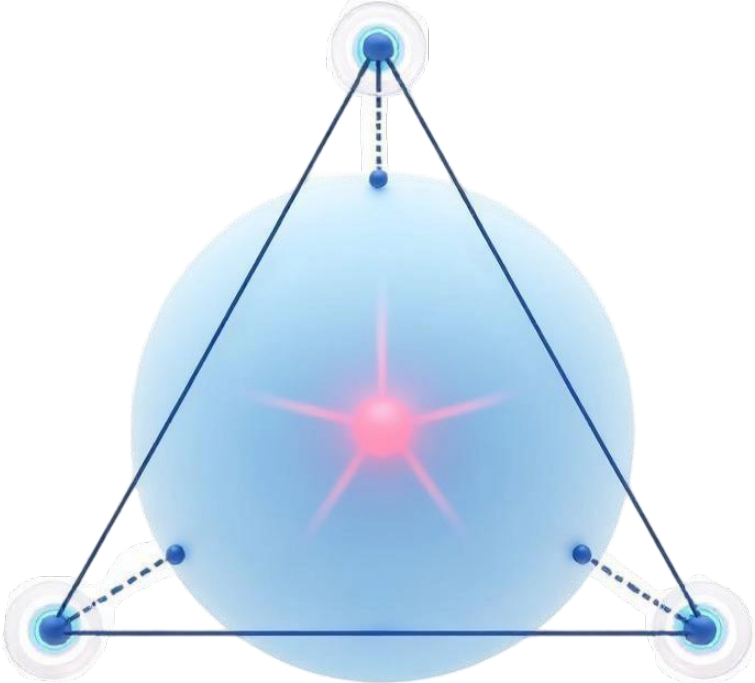} FluidFormer: Transformer with Continuous Convolution for Particle-based Fluid Simulation}
\author{
    Nianyi Wang\textsuperscript{}\equalcontrib,
    Yu Chen\textsuperscript{}\equalcontrib,
    Shuai Zheng\textsuperscript{}\thanks{Corresponding author: shuaizheng@xjtu.edu.cn}
}
\affiliations{
    \textsuperscript{}School of Software Engineering, Xi’an Jiaotong University, Xi’an, 710049, China
}

\usepackage{bibentry}
\begin{document}
\begin{sloppypar}
\maketitle

\begin{abstract}
Learning-based fluid simulation networks have been proven as viable alternatives to traditional numerical solvers for the Navier-Stokes equations. Existing neural methods follow Smoothed Particle Hydrodynamics (SPH) frameworks, which inherently rely only on local inter-particle interactions. However, we emphasize that global context integration is also essential for learning-based methods to stabilize complex fluid simulations. We propose the first Fluid Attention Block (FAB) with a local-global hierarchy, where continuous convolutions extract local features while self-attention captures global dependencies. This fusion suppresses the error accumulation and models long-range physical phenomena. Furthermore, we pioneer the first Transformer architecture specifically designed for continuous fluid simulation, seamlessly integrated within a dual-pipeline architecture. Our method establishes a new paradigm for neural fluid simulation by unifying convolution-based local features with attention-based global context modeling. FluidFormer demonstrates state-of-the-art performance, with stronger stability in complex fluid scenarios.
\end{abstract}

\section{Introduction}
Fluid simulation has recently emerged as a promising area for machine learning. Given the high computational cost of traditional Navier-Stokes equation solvers, deep learning methods are increasingly being developed to replace conventional physics-based approaches. A common representation is to model fluids as dense 3D point clouds, characterized by position and velocity vectors \cite{ummenhofer2019lagrangian}. By calculating forces between particles, we can predict the particle states in subsequent frames. This particle-based fluid simulation is formally known as Smoothed Particle Hydrodynamics (SPH) \cite{ye2019smoothed, liu2010smoothed}.

In the SPH framework, the fluid properties at arbitrary spatial points are computed by kernel-weighted averaging of attributes from neighboring particles. This kernel assigns distance-based weights to particles within a finite radius, with values vanishing outside this range. Theoretically, restricting computation to local particle interactions aligns with the fundamental physics of fluid dynamics, which existing fluid networks universally follow \cite{chen2024dualfluidnet,shao2022transformer,prantl2022guaranteed}. However, local-only computation in neural network methods will induce computational instability. This occurs because the local computation via convolution kernels introduces errors that propagate as long-range inaccuracies through fluid-mediated interactions, resulting in accumulated systematic deviations.

In this paper, we demonstrate that global feature integration is essential for learning-based methods, enabling them to capture long-range physical phenomena and enhancing stabilization in complex fluid simulations. We propose the first Fluid Attention Block (FAB), a novel local-global hierarchical architecture that integrates continuous convolutions for local feature extraction and self-attention for comprehensive long-range dependency modeling. Besides, to adapt Transformer for 3D fluid particle simulation, we introduce 3D Rotary Position Encoding (3D-RoPE) and Type-aware Embedding specifically designed for fluid particles. FluidFormer employs dual-pipe'li'ne architecture combining Main Path with Physics-guided Path to balance fluid dynamics  capturing and physical laws adherence.

We conducted comprehensive experiments on the classic water dataset \cite{ummenhofer2019lagrangian} and the complex Fueltank dataset \cite{chen2025pioneering, zheng2021topology}, which is characterized by scene complexity and dynamic intensity. Experiments demonstrate FluidFormer's state-of-the-art performance across multiple datasets, with superior generalization capabilities and stability, especially in complex scenarios.

In general, the main contributions of this paper include:
\begin{itemize}
\item
Diverging from the prevailing consensus in SPH-based neural networks, we demonstrate that global context integration, rather than local-only computations, is essential for learning-based models to stabilize fluid simulation.
\item 
We propose the first Transformer architecture designed for continuous fluid simulation, with domain-specific innovations such as 3D-RoPE for particles, Type-aware Embedding, and Local-global Fluid Attention Block.
\item
We present a new paradigm for neural fluid simulation that integrates Transformer within a dual-pipeline architecture. This design achieves optimal balance between learning stability and adherence to physical laws, demonstrating across-the-board performance supremacy.
\end{itemize}

\section{Related Work}
\subsection{Learning-based Fluid Particle Simulation}
Recent studies predict fluid particle states via deep neural networks by extracting features from neighboring particles \cite{SAHA2021359, morton2018deep, tompson2017accelerating, ling2016reynolds}. Two dominant SPH-inspired neural methods are as follows:
\subsubsection{Graph-based Methods}
Graph-based Methods represent fluid particles as nodes and their interactions as edges \cite{shao2022transformer,sanchez2020learning,li2018learning,battaglia2016interaction}. However, this discretization compromises fluid continuum properties. The explicit dynamic graph incurs computational overhead.
\subsubsection{Continuous-Convolution Methods}
Continuous convolutions (CConv) inherently preserve physical continuity required by Navier-Stokes equations, aggregating neighbor features through differentiable convolution kernels \cite{ummenhofer2019lagrangian}. ASCC \cite{prantl2022guaranteed} incorporates antisymmetric kernel designs in CConv, enforcing strong momentum conservation constraints. PioneerNet and DualFluidNet \cite{chen2025pioneering,chen2024dualfluidnet} achieves optimal balance between CConv and ASCC through multi-path network architectures. We build upon the continuous convolution approaches, incorporating insights from validated multi-pipeline architectures to preserve core fluid modeling capabilities.

\subsection{Local and Global Fluid Features}
Existing fluid networks focus only on local features, following the prevailing consensus of SPH. Methods \cite{chen2024dualfluidnet, prantl2022guaranteed, ummenhofer2019lagrangian} aggregate neighbor attributes within radius \(R\) by 3D spherical kernels, with kernel values vanishing beyond \(R\). PioneerNet \cite{chen2025pioneering} introduced fully-connected (FC) layers within its multi-pipeline architecture to broadly control the overall fluid motion within a reasonable range. However, FC layers exhibit limited representational capacity for capturing complex global contexts. More critically, it overlooks the greater potential of global context modeling, failing to recognize its fundamental necessity in fluid neural networks. Our work bridges this gap through a local-global hierarchical Fluid Attention, where continuous convolutions encode local features while self-attention captures global dependencies. 

\subsection{Attention for Fluid Simulation}
Transformers demonstrate exceptional long-range contextual modeling capabilities in NLP \cite{vaswani2017attention} and Computer Vision\cite{khan2022transformers, dosovitskiy2020image}. Existing efforts like DualFluidNet and PioneerNet \cite{chen2024dualfluidnet, chen2025pioneering} employ simplistic soft-attention for feature fusion. Although TIE \cite{shao2022transformer} attempted to integrate Transformers into graph-based fluid simulation, its attention mechanism remains confined to local radius \(R\) neighborhoods. In addition, it still suffers from the inherent limitations of graph-based constraints. To overcome these limitations, we propose the first Transformer with Continuous Convolution architecture specifically designed for fluid simulation. Furthermore, to mitigate the quadratic memory growth of attention computation with increasing particle counts, we use Flash Attention \cite{dao2022flashattention} to reduce GPU memory overhead while maintaining exact attention accuracy.

\section{Problem Formulation}\label{section: Problem Formulation}
We formulate fluid simulation within a SPH framework, extending Position-Based Fluids (PBF) through neural feature-driven dynamics \cite{macklin2013position}. Consider a discrete system comprising two disjoint particle sets:
\begin{itemize}
    % \item \text{Fluid particles} $\{\phi_i^n \mid \phi_i^n = (\mathbf{x}_i^n, f_i=[ 1, \mathbf{v}_i^n, \nu_i])\}$, $ i \in [1, N] $
    % \item \text{Boundary particles} $\{\varphi_j \mid \varphi_j = (\mathbf{x}_j, n_j)\}$, $ j \in [1, M] $
    \item Fluid particles $\{\phi_i^n \mid \phi_i^n = (\mathbf{x}_i^n, f_i = [1, \mathbf{v}_i^n, u_i^n]),\ i \in [1, N]\}$
    \item Boundary particles $\{\varphi_j \mid \varphi_j = (\mathbf{x}_j, n_j),\ j \in [1, M]\}$
\end{itemize}
At timestep $n$, each fluid particle $\phi_i^n$ is represented by a tuple containing its position $\mathbf{x}_i^n$, velocity $\mathbf{v}_i^n$, and viscosity coefficient $\nu_i$ as feature vectors $f_i$. Similarly, each boundary particle $\psi_j$ is defined by a tuple containing its position $\mathbf{x}_j$ and surface normal $n_j$. 

We first compute intermediate states driven by external forces $\mathbf{F}_{ext}$ via Heun's predictor-corrector scheme:
\begin{equation}
    \tilde{\mathbf{v}}_i^n = \mathbf{v}_i^n + \Delta t \frac{\mathbf{F}_{ext}}{m_i}, 
    \label{eq:predictor_v}
\end{equation}
\begin{equation}
    \tilde{\mathbf{x}}_i^n = \mathbf{x}_i^n + \Delta t \frac{\mathbf{v}_i^n + \tilde{\mathbf{v}}_i^n}{2}.
    \label{eq:predictor_x} 
\end{equation}
However, the displacement caused by inter-particle forces cannot be directly computed by simple formulas and must be inferred through implicit physical relationships. We use a neural network $\mathcal{G}_\theta$ to predict position corrections $\Delta \mathbf{x}_i$:
\begin{equation}
    [\Delta \mathbf{x}_1, \ldots, \Delta \mathbf{x}_N] = \mathcal{G}_\theta \left( \{ \phi_1^n, \ldots, \phi_N^n \}, \{ \varphi_1, \ldots, \varphi_M \} \right).
    \label{eq:neural_correction}
\end{equation}

Our ultimate objective is to predict the particle state at timestep $n+1$ by updating:
\begin{equation}
    \mathbf{x}_i^{n+1} = \tilde{\mathbf{x}}_i^n + \Delta \mathbf{x}_i, \label{eq:update_x}
\end{equation}
\begin{equation}
    \mathbf{v}_i^{n+1} = \frac{(\mathbf{x}_i^{n+1} - \mathbf{x}_i^n)}{\Delta t}. \label{eq:update_v}
\end{equation}

\section{Method}
\subsection{3D Rotary Position Encoding (3D-RoPE)}
For 3D fluid particles, the coordinates explicitly encode spatial locations, but their low-level vector representations lack inherent modeling of implicit geometric relationships. In large language models (LLMs) and vision-language models (VLMs), rotary position embedding (RoPE) is commonly employed to encode implicit relative positional dependencies by applying rotation matrices to query and key vectors in the attention mechanism. This technique is applied to 1D sequences such as text \cite{su2024roformer}, with adaptations that extend it to 2D data such as images \cite{heo2024rotary, wang2024qwen2}. We introduce 3D-RoPE and integrate it into particle-based fluid simulations. This extension encodes spatial relationships among fluid particles, enhancing the modeling of long-range hydrodynamic interactions.

\begin{figure}
\centering
\begin{subfigure}[b]{0.58\linewidth}
    \centering
    \includegraphics[scale=0.75]{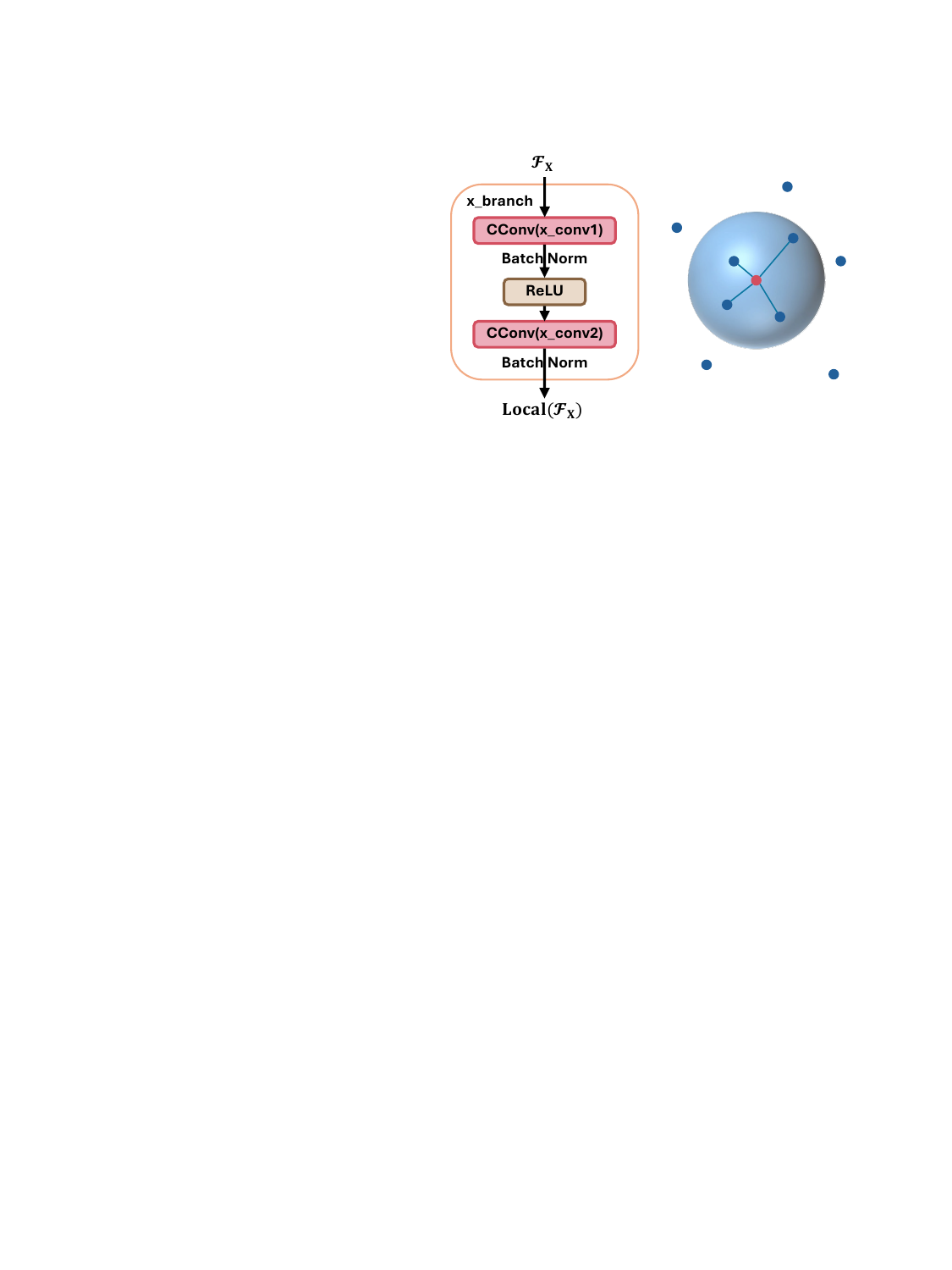}
    \caption{Architecture}
    \label{fig:local_architecture}
\end{subfigure}
\hfill
\begin{subfigure}[b]{0.4\linewidth}
    \centering
    \includegraphics[width=\linewidth]{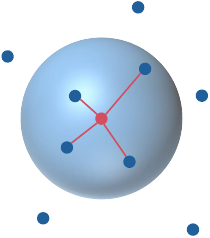}
    \caption{Visualization}
    \label{fig:local_sphere}
\end{subfigure}
\caption{The Local Feature Extractor based on the CConv kernel. The visualization demonstrates that it performs convolutional computations exclusively on neighbor particles within a specific range of each target particle.}
\label{fig:local}
\end{figure}

\begin{figure}[t]
    \centering
    \begin{subfigure}[b]{0.58\linewidth}
        \centering
        \includegraphics[scale=0.75]{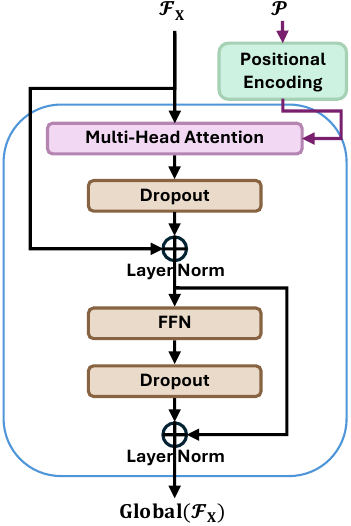}
        \caption{Architecture}
        \label{fig:global_architecture}
    \end{subfigure}
    \hfill
    \begin{subfigure}[b]{0.4\linewidth}
        \centering
        \includegraphics[width=\linewidth]{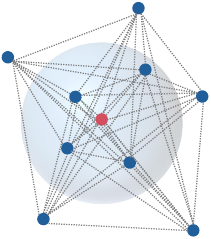}
        \caption{Visualization}
        \label{fig:global_sphere}
    \end{subfigure}
    \caption{The Global Feature Extractor establishes global dependencies for particle state propagation through the Multi-Head Attention mechanism on all particles.}
    \label{fig:global}
\end{figure}

Specifically, we represent the position of the 3D particle \(i\) as $\mathbf{x}_i=(x,y,z)$. 
In 3D-RoPE, the rotation angle \(\theta\) for each dimension pair \((2k,2k+1)\) is computed as:
\begin{equation}
\theta = b^{-2k/d}, \quad k=0,1,\ldots,\lfloor d/2 \rfloor -1.
\label{eq: theta}
\end{equation}
Here, $b$ is a hyperparameter set to 10000, $k$ indexes dimension pairs, and $d$ represents the embedding dimension.

The block-diagonal matrix \(\mathbf{R}_{\mathbf{x}_i}\) comprises three rotation matrices \(\mathbf{R}_{x}, \mathbf{R}_{y}, \mathbf{R}_{z}\) on its diagonal, with \(\mathbf{R}_\alpha \!=\! \bigl( \begin{smallmatrix} \cos\alpha\theta & \!\!-\sin\alpha\theta \\ \sin\alpha\theta & \!\!\cos\alpha\theta \end{smallmatrix} \bigr)\) for \(\alpha \in \{x,y,z\}\). $\theta$ is defined in Equation \ref{eq: theta}.

\setlength{\arraycolsep}{1.0pt}
\begin{equation}
    \mathbf{R}_{\mathbf{x}_i} = 
    {\small
    \left( \begin{array}{cc;{2pt/2pt}cc;{2pt/2pt}cc}
    \cos x\theta & -\sin x\theta & 0 & 0 & 0 & 0 \\
    \sin x\theta & \cos x\theta & 0 & 0 & 0 & 0 \\
    \cdashline{1-6}[1pt/2pt]
    0 & 0 & \cos y\theta & -\sin y\theta & 0 & 0 \\
    0 & 0 & \sin y\theta & \cos y\theta & 0 & 0 \\
    \cdashline{1-6}[1pt/2pt]
    0 & 0 & 0 & 0 & \cos z\theta & -\sin z\theta \\
    0 & 0 & 0 & 0 & \sin z\theta & \cos z\theta
    \end{array} \right)
    }
\end{equation}
This rotation mechanism integrates relative distance information between particles into the attention computation, as will be demonstrated in the next subsection \ref{subsection:Fluid_Attention_Block}, and exhibits extrapolation capability to varying particle counts.

\subsection{Fluid Attention Block}\label{subsection:Fluid_Attention_Block}
\subsubsection{Local Feature Extractor}
Traditional discrete convolution fails to effectively model local particle interactions in continuum physical spaces. Building on previous work \cite{chen2025pioneering,ummenhofer2019lagrangian} that demonstrated the efficacy of continuous convolutions in approximating the SPH fluid dynamics kernels, CConv extracts the features of the particle at position $\mathbf{x}$ using the positions of neighboring particles $\mathbf{x}_{i}$ and the feature vector $f_i$ defined in section \ref{section: Problem Formulation}:
\begin{equation}
    \begin{aligned}
\text{CConv}_g&=\left ( f*g \right )\left ( \mathbf{x} \right ) \\   &=\sum_{i\in \mathcal{N}\left (\mathbf{x},R  \right )}^{}a\left (\mathbf{x}_{i}, \mathbf{x}  \right )f_{i}g\left ( \Lambda \left ( \mathbf{x}_{i} - \mathbf{x} \right )  \right ).
    \end{aligned}
\label{Eq: CConv}
\end{equation}
$g$ is the convolution kernel. The spherical neighborhood $\mathcal{N}\left (\mathbf{x},R  \right )$ defines the particle set within radius $R$ of $\mathbf{x}$. Mapping function $\Lambda$ dynamically adapts the kernel shape to non-uniform particle distributions. Window function $a\left (\mathbf{x}_{i}, \mathbf{x}  \right )$ modulates the contribution weights of neighboring particles.

Denote the input particle features as $\mathcal{F}_{\mathbf{X}}$, \textit{Local Feature Extractor} module is defined as:
\begin{equation}
    \text{Local}(\mathcal{F}_{\mathbf{X}}) = \text{BN}\Big(\text{CConv}\big(\text{ReLU}(\text{BN}(\text{CConv}(\mathcal{F}_{\mathbf{X}}))\big)\big).
\end{equation}
This cascade aggregates neighborhood interactions via kernel propagation, enriching local representations in particle simulation. The architecture is shown in Figure \ref{fig:local}.

\begin{figure*}[t]
    \centering
    \includegraphics[width=0.95\textwidth]{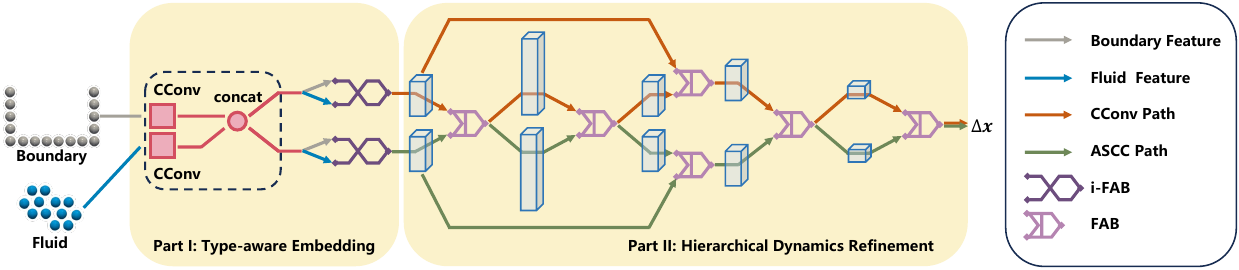}
    \caption{The braided fusion architecture integrates two core components:
(1) Type-aware Embedding generates fused embeddings that intrinsically distinguish fluid and boundary particle types, leveraging CConv operations and i-FAB to model intricate fluid-boundary coupling.
(2) Hierarchical Dynamic Refinement employs a dual-pipeline framework balancing fluid modeling with physical constraints, where the Fluid Attention Block (FAB) holistically captures local-global features across multiple scales.}
    \label{fig:framework}
\end{figure*}
\begin{figure}[t]
  \centering
  \begin{subfigure}{0.53\columnwidth}
    \centering
    \includegraphics[scale=0.75]{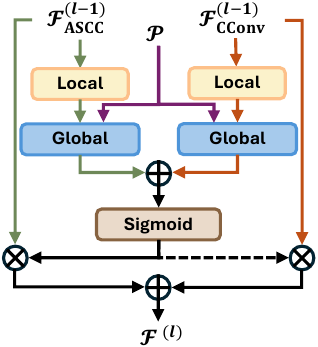}
    \caption{FAB}
    \label{fig:FAB}
  \end{subfigure}
  \hfill 
  \begin{subfigure}{0.45\columnwidth}
    \centering
    \includegraphics[scale=0.75]{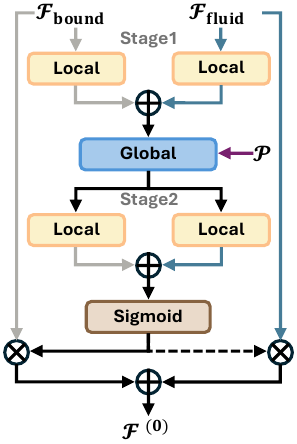}
    \caption{i-FAB}
    \label{fig:i-FAB}
  \end{subfigure}
  \vspace{-0.2cm} 
  \caption{Architectures of two types of Fluid Attention Block (FAB and i-FAB). FAB employs a local-to-global hierarchical structure to fuse multiscale features from dual pathways, while i-FAB adopts a two-stage iterative architecture for enhanced modeling of fluid-boundary coupling.}
  \label{fig:2FAB}
\end{figure}
\subsubsection{Global Feature Extractor}

While traditional SPH methods depend solely on local particle interactions, purely convolutional neural networks induce computational instability. This occurs as convolution-kernel computations introduce errors that propagate into long-range inaccuracies through fluid interactions, causing systemic error accumulation and global instability, particularly in scenarios involving violent fluid motion. To address this, we introduce a \textit{Global Feature Extractor}  that explicitly models global long-range dependencies to maintain fluid stability. We first define the attention score between particles $i$ and $j$ with 3D-RoPE:
\begin{equation}
\begin{aligned}
    \text{Attention}(i, j) &= \operatorname{softmax}\left( \frac{ (\mathbf{R}_{\mathbf{x}_i} \mathbf{q}_i)^T (\mathbf{R}_{\mathbf{x}_j} \mathbf{k}_j) }{ \sqrt{d} } \right) \mathbf{v}_j \\
    &= \operatorname{softmax}\left( \frac{ \mathbf{q}_i^T \mathbf{R}_{\mathbf{x}_i}^T \mathbf{R}_{\mathbf{x}_j} \mathbf{k}_j }{ \sqrt{d} } \right) \mathbf{v}_j\\
    &= \operatorname{softmax}\left( \frac{ \mathbf{q}_i^T \mathbf{R}_{\mathbf{x}_j-\mathbf{x}_i} \mathbf{k}_j }{ \sqrt{d} } \right) \mathbf{v}_j.
\end{aligned}
\end{equation}
The 3D-RoPE transforms the attention computation by rotating query/key vectors with a rotation matrix $\mathbf{R}_{\mathbf{x}_j-\mathbf{x}_i}$. This diagonal rotation matrix maintains vector norms while explicitly embedding spatial relationships based on the relative displacement vector between particles $i$ and $j$. 

Then Multi-Head Attention (MHA) is computed as follows, consistent with previous work \cite{vaswani2017attention}:
\begin{equation}
    \text{MHA}(\mathbf{Q},\mathbf{K},\mathbf{V}) = \text{Concat}(\mathbf{head}_{(1)},\dots,\mathbf{head}_{(h)})\mathbf{W}^O.
\end{equation}
For the $m^{th}$ head ($1\leq m \leq h=4$):
\begin{equation}
    \mathbf{head}_{(m)} = \text{Attention}(\mathbf{Q}\mathbf{W}_{(m)}^Q,\mathbf{K}\mathbf{W}_{(m)}^K,\mathbf{V}\mathbf{W}_{(m)}^V).
\end{equation}
As illustrated in Figure \ref{fig:global}, this extractor leverages the all-to-all connectivity of Multi-Head Attention to implicitly construct a continuous geometry-aware interaction topology without discrete graph constraints. It captures dependencies across the entire fluid domain, minimizing error accumulation in inter-region propagation through truly global information exchange. Additionally, we optimize the attention computation using Flash Attention \cite{dao2022flashattention}, significantly reducing both computational and memory overhead.

\subsubsection{Local-global Hierarchy}
Learning-based fluid simulation must overcome challenges arising from particle system disorder, unstructured configurations, and multiscale physical features. Building upon our feature extractors, we propose the \textit{Fluid Attention Block} (\textit{FAB}) with dedicated local-global hierarchical architecture. Serial network architectures tend to form information bottlenecks during feature propagation, causing feature attenuation and loss of critical physical details. \textit{FAB} dynamically performs dual feature extraction through adaptive soft-attention mechanisms, as illustrated in Figure \ref{fig:FAB}. For the input features $\mathcal{F}_{\mathbf{X}}$ and $\mathcal{F}_{\mathbf{Y}}$, \textit{FAB} separately processes them through local and global feature extractors, followed by feature fusion via soft-attention:
\begin{equation}
    \mathcal{F} = \Gamma\left(\mathcal{F}_{\mathbf{X}}, \mathcal{F}_{\mathbf{Y}} \right),
    \label{eq:fusion}
\end{equation}
\begin{equation}
    \Gamma\left(\mathcal{F}_{\mathbf{X}}, \mathcal{F}_{\mathbf{Y}} \right) = \gamma \big( \mathcal{F}_{\mathbf{X}} \otimes \sigma(\mathcal{F}_{\text{fused}}) + \mathcal{F}_{\mathbf{Y}} \otimes (1-\sigma(\mathcal{F}_{\text{fused}})) \big),
\end{equation}
\begin{equation}
    \mathcal{F}_{\text{fused}} = \text{Global}_x(\text{Local}_x(\mathcal{F}_{\mathbf{X}})) \oplus \text{Global}_y(\text{Local}_y(\mathcal{F}_{\mathbf{Y}})).
\end{equation}
Here, $\sigma$ denotes the sigmoid function, and the scaling parameter $\gamma$ is set to amplify feature disparities.

\subsection{Network Architecture}
\subsubsection{Part I: Type-aware Embedding}
To strengthen fluid-solid coupling, we evolve the \textit{FAB} module into an iterative architecture: the \textit{iterative Fluid Attention Block} (\textit{i-FAB}, Figure \ref{fig:i-FAB}). Its inputs $\mathcal{F}_{\text{fluid}}$ and $\mathcal{F}_{\text{bound}}$ represent high-dimensional fluid features and boundary features from independent CConv operations. We propose the \textit{Type-aware Embedding} $\Gamma_{\text{Type}}(\mathcal{F}_{\text{fluid}}, \mathcal{F}_{\text{bound}})$ based on \textit{i-FAB} enables semantic separation and cross-domain interaction between particle types, as shown in Figure \ref{fig:framework}. Deployed in early input stages, \textit{i-FAB} significantly enhances simulation accuracy.

\setlength{\tabcolsep}{3mm}
\begin{table*}[!t]
  \centering
    \begin{tabular}{lcccccccc}
      \toprule[0.6pt]
      \multirow{2}{*}{Method} & \multicolumn{2}{c}{CD (\(mm\))} & \multicolumn{2}{c}{EMD (\(mm\))} & \multirow{2}{*}{\shortstack{$n$-frame Sequence Error\\ (\(mm\))}} & \multirow{2}{*}{\shortstack{Max Density Error\\  (\(g/cm^{3}\))}} & \multirow{2}{*}{Time (s)} \\ \cline{2-3} \cline{4-5}
                              & t+1 & t+2 & t+1 & t+2 & & & \\ \midrule \midrule
      CConv                & 0.709 & 2.093 & 0.129 & 0.294 & 33.697 & 0.189 & \textbf{0.019}\\
      DMCF               & 0.718 & 2.171 & 0.123 & 0.249 & 34.753 & 0.104 & 0.098\\
      TIE               & 0.723 & 2.312 & 0.136 & 0.313 & 35.542 & 0.125 & 0.142\\
      DualFluidNet        & 0.541 & 1.504 & 0.120 & 0.227 & 30.982 & 0.079 & 0.051\\
      PioneerNet        & 0.520 & 1.454 & 0.113 & 0.210 & 29.583 & 0.075 & 0.048\\
      Ours                   & \textbf{0.418} & \textbf{1.152} & \textbf{0.099} & \textbf{0.194} & \textbf{27.861} & \textbf{0.068} & 0.057 \\ \bottomrule[0.6pt]
    \end{tabular}
  \caption{Quantitative comparison on Liquid3D(complex). Evaluations on this classic water dataset quantify fundamental fluid modeling capabilities, revealing our method's superior performance even on low-variance data.}
  \label{tab:Fluid_experiments}
\end{table*}

\begin{figure*}[!t]
\centering
    \includegraphics[width=0.95\linewidth]{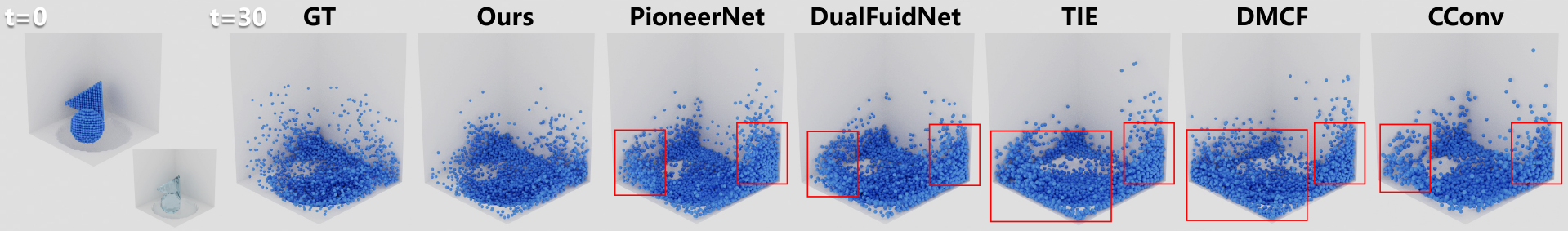}
    \caption{Qualitative comparison on Liquid3D(complex). Splash morphology after two differently shaped liquid blocks impact a circular groove. Our method achieves ground-truth comparable visual fidelity.}
    \label{fig:liquid}
\end{figure*}

\subsubsection{Part II: Hierarchical Dynamics Refinement}
We integrate Transformer within a dual-pipeline framework, striking an optimal balance between learning stability and physical constraints. Figure \ref{fig:framework} depicts the complete architecture. The upper pathway's Global Feature Extractor is based on \textit{CConv} \cite{ummenhofer2019lagrangian} (Eq. \ref{Eq: CConv}). To strengthen physical constraint modeling, we introduce the Antisymmetric Continuous Convolution (\textit{ASCC}) and base the Local Feature Extractor in the lower pathway on it. \textit{ASCC} is defineds as:
\begin{equation}
    \begin{aligned}
    \text{ASCC}_{g_s}&=\left ( f*g_s \right )\left ( \mathbf{x} \right ) \\ &=\sum_{i\in \mathcal{N}\left (\mathbf{x},R  \right )}^{}a\left (\mathbf{x}_{i}, \mathbf{x}  \right )(f+f_{i})g_s\left ( \Lambda \left ( \mathbf{x}_{i} -\mathbf{x} \right )  \right ).
    \end{aligned}
\end{equation}
\textit{ASCC} is a variant of \textit{CConv} that halves the convolution kernel $g$ and applies mirrored parameters with sign inversion to get the $g_s$. \cite{prantl2022guaranteed, chen2024dualfluidnet} prove rigorously that this antisymmetric design can introduce strong physical constraints into continuous convolution, ensuring strict adherence to momentum conservation laws.

After the Type-aware Embedding Module calculates $\mathcal{F}^{(0)} = \Gamma_{\text{Type}}(\mathcal{F}_{\text{fluid}}, \mathcal{F}_{\text{bound}})$, the computational workflow of Hierarchical Dynamics Refinement with residual connection $\mathcal{F}_{\text{Res}}$ can be formulated as:
\begin{equation}
    \mathcal{F}^{(l-1)}_{\text{CConv}}, \mathcal{F}^{(l-1)}_{\text{ASCC}}=\mathcal{F}^{(l-1)}.
\end{equation}
\begin{multline}
    \mathcal{F}^{(l)} = \Gamma^{(l)}\left( 
      \underbrace{\Psi_{\text{CConv}}^{(l)}(\mathcal{F}^{(l-1)}_{\text{CConv}})}_{\text{CConv Path}},
      \underbrace{\Psi_{\text{ASCC}}^{(l)}(\mathcal{F}^{(l-1)}_{\text{ASCC}})}_{\text{ASCC Path}}, 
      \mathcal{F}_{\text{Res}}^{(l-2)}
    \right),\\
    \text{for } l = 1,2,3,4 \quad \text{with} \quad \mathcal{F}_{\text{Res}}^{(-1)} = \mathcal{F}_{\text{Res}}^{(0)} = \emptyset.
\end{multline}
\begin{equation}
    \Delta \mathbf{x} = \frac{1}{\kappa} \mathbf{W}_{\text{out}} \mathcal{F}^{(4)}, \quad \mathbf{W}_{\text{out}} \in \mathbb{R}^{3 \times C}.
\end{equation}
Finally, we obtain the particle position offset $\Delta \mathbf{x}$ driven by inter-particle forces, as described in Section \ref{section: Problem Formulation}. $\Psi_{\text{CConv}}$ and $\Psi_{\text{ASCC}}$ represent convolution layers based on CConv and ASCC respectively. The scaling factor $\kappa$ is set to 128.

Breaking through the local-only paradigm of traditional SPH methods, our proposed FluidFormer establishes a new paradigm for neural fluid simulation  that effectively integrates multi-level local-global features for neural fluid simulation, achieving enhanced stability in fluid simulations.

\section{Experiments}
\subsection{Experimental Setup}
\subsubsection{Datasets}

We employ the Liquid3D benchmark dataset \cite{ummenhofer2019lagrangian,bender2015divergence}, which simulates the trajectory of the fluid block falling inside the basic geometric container, providing standardized scenarios for evaluating the fundamental simulation capabilities of fluids.

Furthermore, we evaluate on the more complex benchmark Fueltank \cite{chen2025pioneering}, which simulates fuel sloshing in intricate tank structures under random aircraft-induced perturbations. This critically tests model accuracy and robustness in highly dynamic, violent fluid regimes.

\subsubsection{Evaluation Metrics}
We evaluate spatial accuracy using Chamfer Distance (CD) for particle set matching and Earth Mover's Distance (EMD) for distribution similarity. Short-term predictive capability is assessed via two-frame trajectory forecasts. For long-term error accumulation, we employ $n$-frame Sequence Error ($n$-SE). Additionally, Maximum Density Error (MDE) quantifies adherence to incomp physical constraints. Single-frame inference latency is measured to benchmark computational efficiency.

\setlength{\tabcolsep}{3mm}
\begin{table*}[t]
  \centering
    \begin{tabular}{lcccccccc}
      \toprule[0.6pt]
      \multirow{2}{*}{Method} & \multicolumn{2}{c}{CD (\(mm\))} & \multicolumn{2}{c}{EMD (\(mm\))} & \multirow{2}{*}{\shortstack{$n$-frame Sequence Error\\ (\(mm\))}} & \multirow{2}{*}{\shortstack{Max Density Error\\  (\(g/cm^{3}\))}} & \multirow{2}{*}{Time (s)} \\ \cline{2-3} \cline{4-5}
                              & t+1 & t+2 & t+1 & t+2 & & & \\ \midrule \midrule
      CConv                & 1.713 & 4.110 & 0.607 & 1.002 & 166.205 & 0.175 & \textbf{0.026}\\
      DMCF               & 1.532 & 3.985 & 0.263 & 0.424 & 132.263 & 0.051 & 0.653\\
      TIE               & 1.695 & 4.002 & 0.278 & 0.548 & 142.267 & 0.098 & 1.475\\
      DualFluidNet        & 1.403 & 3.977 & 0.237 & 0.507 & 41.013 & 0.018 & 0.213\\
      PioneerNet        & 1.322 & 3.507 & 0.206 & 0.427 & 36.307 & 0.014 & 0.185\\
      Ours                   & \textbf{1.012} & \textbf{2.481} & \textbf{0.132} & \textbf{0.230} & \textbf{24.442} & \textbf{0.008} & 0.226 \\ \bottomrule[0.6pt]
    \end{tabular}
  \caption{Quantitative comparison on Tank I. Comparative evaluation in such complex scenarios critically highlights the differences between models in terms of long-term stability and robustness. Our proposed FluidFormer demonstrates its stability advantage particularly under violent fluid motion or intricate boundaries.}
  \label{tab:Fuel_experiments}
\end{table*}

\subsubsection{Implementation Details}
All models were implemented in PyTorch and trained on NVIDIA A800 GPUs. We employed the Adam optimizer ($\beta_1=0.9$, $\beta_2=0.999$) with L2 regularization (weight decay=0.001). The learning rate followed adaptive step decay: initialized at 0.01, then halved at 15k, 25k, 35k, 45k, 50k, and 55k iterations (60k total).

For enhanced temporal stability, we calculate the composite loss as the weighted sum of the two future time steps:
\begin{equation}
\mathcal{L} = \mathcal{L}_{t+1} + \mathcal{L}_{t+2}.
\end{equation}
Each frame loss uses neighbor-aware adaptive weighting:
\begin{equation}
\mathcal{L}_{t+k} = \frac{1}{N} \sum_{i=1}^{N} \left[ \exp\left(-\frac{c_i}{c}\right) \cdot \left\| \hat{\mathbf{x}}_i^{(t+k)} - \mathbf{x}_i^{(t+k)} \right\|_2^{\gamma} \right].
\end{equation}
where $c_i$ represents the fluid neighbor count for particle $i$. $c$ denotes the average neighbor count, set to 40. The exponential term dynamically scales loss by upweighting sparse neighborhoods to enhance interface prediction accuracy and downweighting dense neighborhoods to prevent overfitting, with loss exponent $\gamma=0.5$.

\begin{figure}[t]
\centering
    \includegraphics[width=0.95\linewidth]{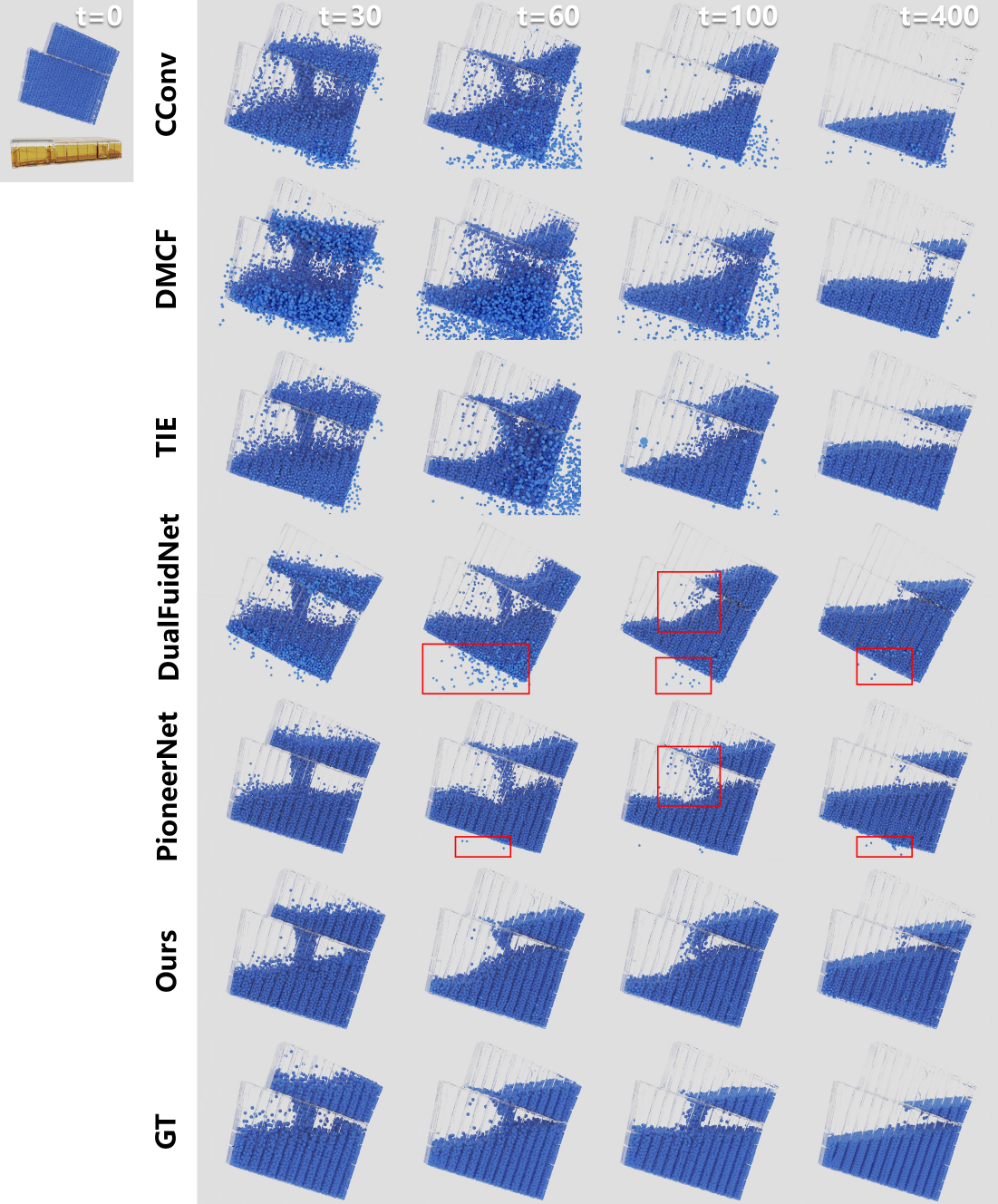}
    \caption{Qualitative comparison on Tank I. Previous methods exhibit severe simulation collapse and unphysical phenomena—the latter are highlighted by red boxes.}
    \label{fig:tank_total}
\end{figure}

\subsection{Comparative Experiments}
\subsubsection{Liquid3D(complex) Dataset}
Table \ref{tab:Fluid_experiments} and Figure \ref{fig:liquid} present evaluations of different methods on this classic water dataset. TIE \cite{shao2022transformer} reveals significant computational overhead and dynamic adaptability bottlenecks caused by fixed-radius neighbor searches on graph structures. While DMCF \cite{prantl2022guaranteed} better adheres to physical constraints than CConv \cite{ummenhofer2019lagrangian}, its forced correction approach via ASCC layer before the output compromises CConv’s inherent continuous expressive capability. This highlights the advantage of our multi-pipeline architecture in balancing these trade-offs. Besides, benefiting from our local-global hierarchy, our method eradicates unphysical particle phenomena observed in baseline simulations: dense particle clustering at container boundaries accompanied by excessive splashing in corners.

\begin{figure}[t]
\centering
    \includegraphics[width=0.95\linewidth]{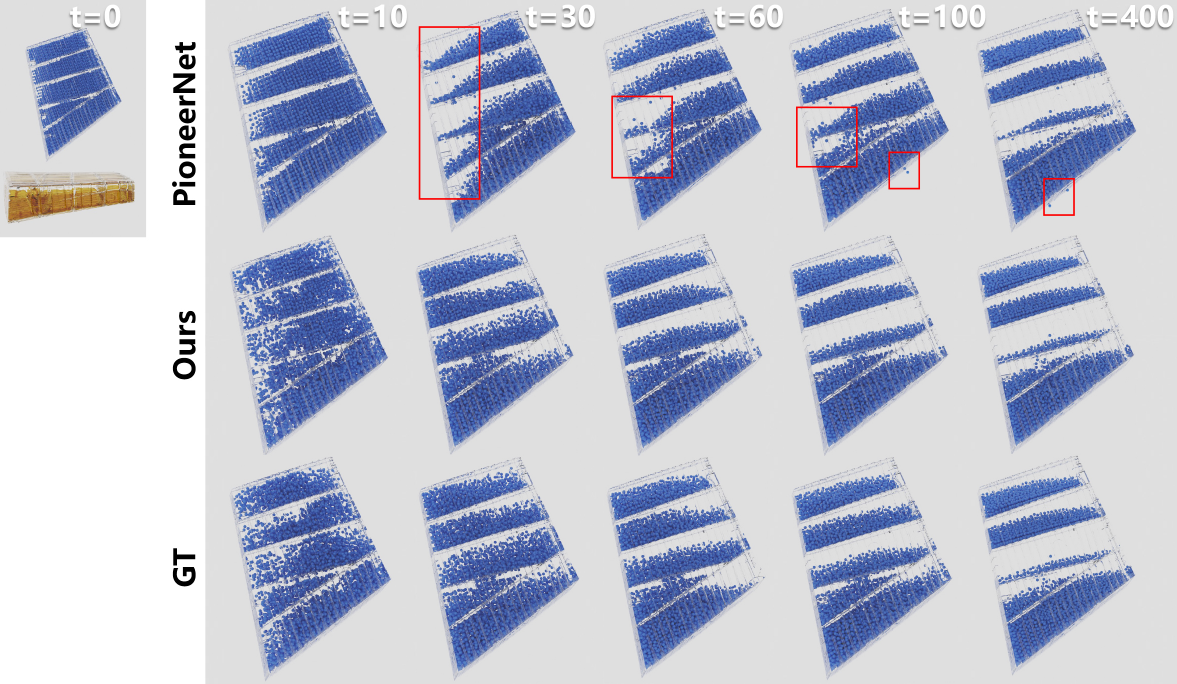}
    \caption{Qualitative comparison on Tank II. PioneerNet exhibits unphysical fluid centroid drift—a critical failure in fluid simulation, while our method is consistent with ground truth.}
  \label{fig:2tank}
\end{figure}

\setlength{\tabcolsep}{1.5mm}
\begin{table}[!t]
  \centering
    \begin{tabular}{lccccc}
      \toprule[0.6pt]
      \multirow{2}{*}{\shortstack{Fueltank\\ Type}} & \multirow{2}{*}{Method} & $\text{EMD}_{t+2}$ & $n\text{-SE}$ & MDE\\
      & & (\(mm\)) & (\(mm\)) & (\(g/cm^{3}\)) \\ \midrule \midrule

      \multirow{2}{*}{Tank II} & PioneerNet               & 0.375 & 36.691 & 0.019 \\
                                 & Ours                   & \textbf{0.319} & \textbf{28.026} & \textbf{0.013} \\
      \midrule
      \multirow{2}{*}{Tank III} & PioneerNet               & 0.475 & 32.252 & 0.015 \\
                                & Ours                   & \textbf{0.360} & \textbf{26.967} & \textbf{0.011} \\ \bottomrule[0.6pt]
    \end{tabular}
      \caption{Quantitative comparison on Tank II and III. Compared against the prior SOTA method PioneerNet, our method achieves a significant error reduction across critical metrics.}
  \label{tab:Fuel_otherexperiments}
\end{table}

\subsubsection{Fueltank Dataset}
Comparative experiments in challenging scenarios (Table \ref{tab:Fuel_experiments}, Figure \ref{fig:tank_total}) particularly demonstrate our multiscale feature learning network's simulation stability in violent fluid motion. Due to weak fluid fitting and low physical accuracy, CConv, DMCF and TIE demonstrated significant simulation collapse. While DualFluidNet and PioneerNet also achieved plausible stability through multi-pipeline architectures, they nevertheless exhibited unphysical fluid drifting and boundary leakage – pathologies stemming from accumulated systematic deviations inherent to local-only paradigms.

Our innovative local-global hierarchy enables FluidFormer to unify convolutional local features with attentional global context in a transformer-based dual-pipeline framework, achieving new state-of-the-art accuracy and robustness. Leveraging Transformers' powerful long-range modeling capabilities, FluidFormer achieves exceptional holistic fluid control beyond continuous convolution fitting. This stabilizes simulations, eliminating any unphysical phenomena like spilling, deformation, drifting, or particle accumulation. Further comparative results in other tanks (Table \ref{tab:Fuel_otherexperiments}, Figure \ref{fig:2tank}) conclusively demonstrate this architectural superiority.

\begin{figure}[t]
  \centering
  \begin{subfigure}[b]{0.95\linewidth}
    \centering
    \includegraphics[width=\linewidth]{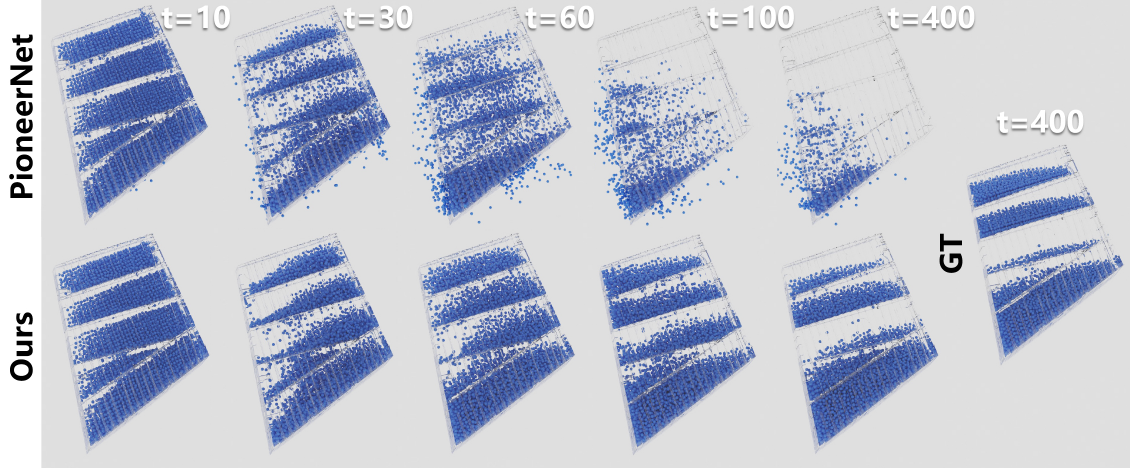}
    \caption{Unseen Tank II}
    \label{fig:unseen_triangle}
  \end{subfigure}
  
  \vspace{0.3cm}
  
  \begin{subfigure}[b]{0.95\linewidth}
    \centering
    \includegraphics[width=\linewidth]{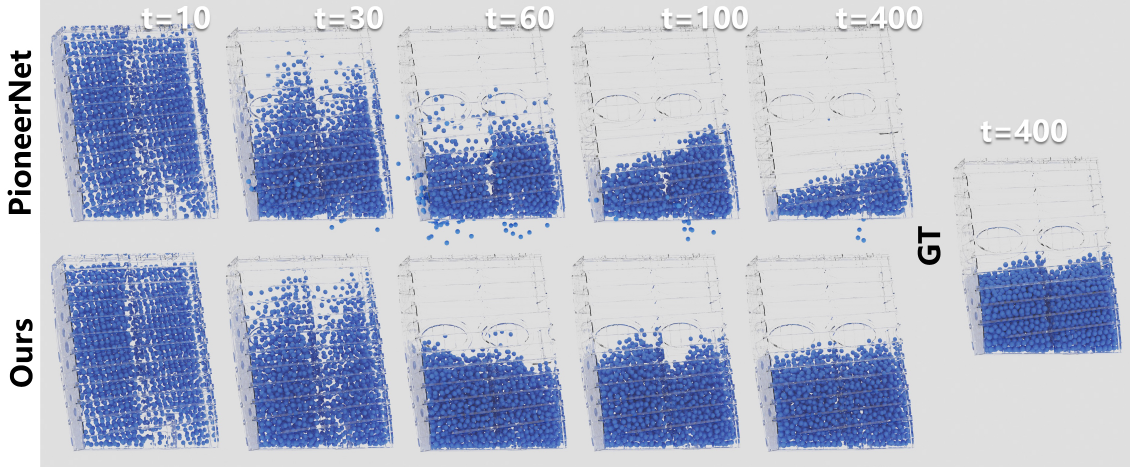}
    \caption{Unseen Tank III}
    \label{fig:unseen_rectangle}
  \end{subfigure}
  
  \caption{Generalization visualization: Simulation results of PioneerNet versus FluidFormer on unseen Tank II and Tank III after training exclusively on Tank I.}
  \label{fig:2tank_unseen}
\end{figure}

\subsection{Generalization in Unseen Fueltank Scenario}
\begin{figure}[t]
\centering
    \includegraphics[width=\linewidth]{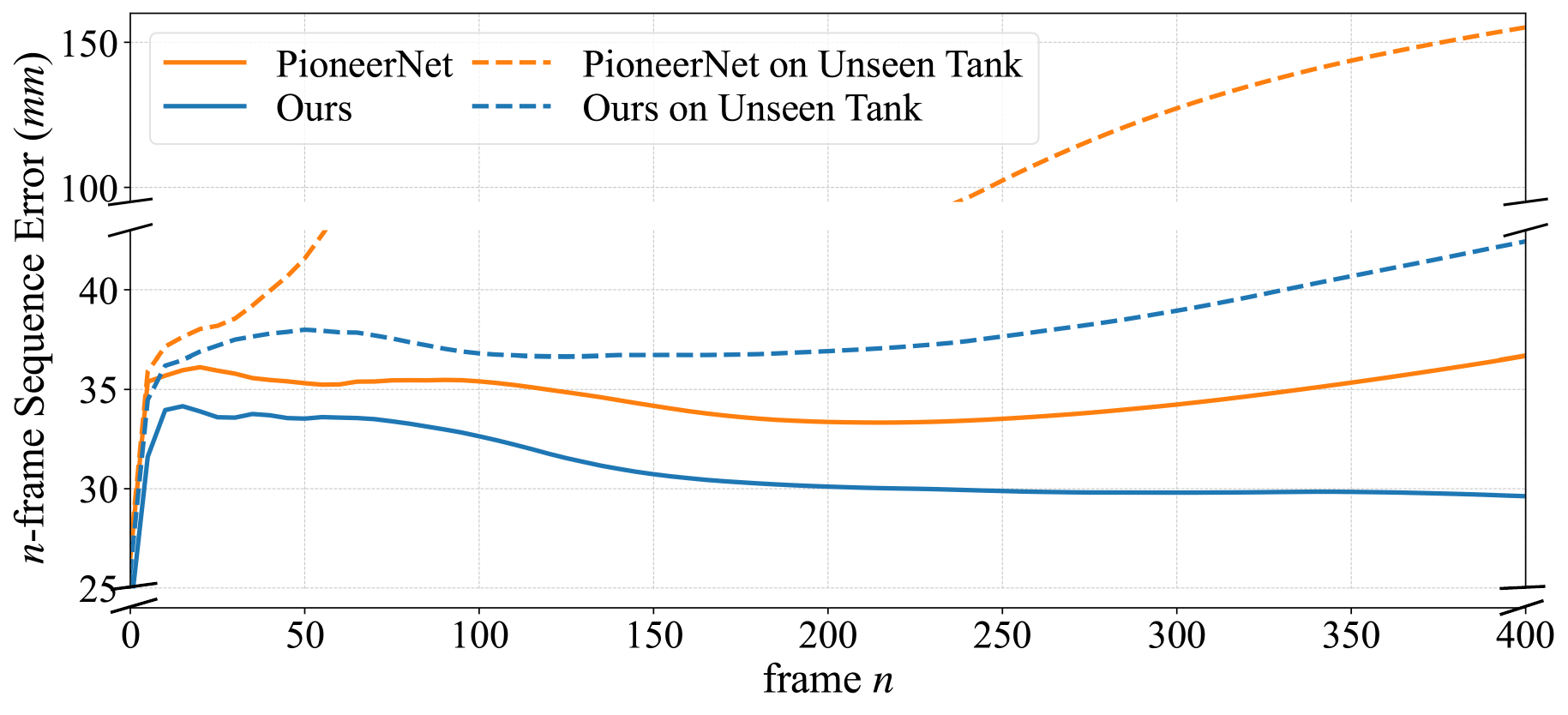}
    \caption{Generalization analysis. Long-term error propagation curves for PioneerNet and FluidFormer on seen vs. unseen Tank II configurations.}
    \label{fig:generalization}
\end{figure}

Previous neural fluid simulators often overfit to specific containment geometries: when trained on one tank shape, they fail to generalize to unseen geometries. To further demonstrate how global attention modeling enhances fluid stability, we compare against PioneerNet (prior SOTA) – both methods trained on identical tank data then tested on completely unseen configurations (Figure \ref{fig:2tank_unseen}). The PioneerNet exhibits severe physical instability, manifesting as particles leakage and unphysical splashing at structural boundaries. In contrast, our approach maintains geometry-invariant boundary awareness, preventing overfitting to specific geometries during training. Figure \ref{fig:generalization} demonstrates our superior generalization capability in unseen scenarios, while PioneerNet suffers catastrophic errors due to overfitting in new scenarios.

% \begin{table}[!t]
%   \centering
%   \begin{tabular}{lccccc}
%     \toprule[0.6pt]
%     Tank & Method & $\text{EMD}$ ($mm$) & $n\text{-SE}$ ($mm$) & MDE ($g/cm^{3}$) \\
%     \midrule \midrule
%     \multirow{2}{*}{Tank I} 
%       & PioneerNet & 0.375 & 36.691 & 0.019 \\
%       & Ours      & \textbf{0.319} & \textbf{28.026} & \textbf{0.013} \\
%     \midrule
%     \multirow{2}{*}{Tank IV} 
%       & PioneerNet & 0.475 & 32.252 & 0.015 \\
%       & Ours      & \textbf{0.360} & \textbf{26.967} & \textbf{0.011} \\ 
%     \bottomrule[0.6pt]
%   \end{tabular}
%   \caption{Quantitative comparison on other fuel tanks.}
%   \label{tab:Fuel_otherexperiments}
% \end{table}

% \begin{table}[!t]
%   \centering
%   \begin{tabular}{lcc}
%     \toprule[0.6pt]
%     Method & $n$-SE ($mm$) & MDE ($g/cm^{3}$) \\
%     \midrule \midrule
%     w/o Global Feature Extractor & 41.024 & 0.019 \\
%     w/o Local Feature Extractor & 75.073 & 0.057 \\
%     w/o Type-aware Embedding & 84.462 & 0.066 \\
%     w/o CConv & 177.151 & 0.141 \\
%     w/o ASCC & 93.524 & 0.073 \\
%     w/o 3D-RoPE & 29.131 & 0.017 \\
%     Ours & \textbf{24.442} & \textbf{0.008} \\
%     \bottomrule[0.6pt]
%   \end{tabular}
%   \caption{Ablation study. Quantitative results demonstrating contributions of local and global feature extractor, and integration of local-global synergy in FluidFormer and its type-aware embedding.}
%   \label{tab:Ablation_Study}
% \end{table}

\subsection{Ablation Studies}
Ablation studies (Table \ref{tab:Ablation_Study}) systematically evaluate contributions of FluidFormer's core components. CConv's continuous modeling and ASCC's physical constraints constitute indispensable foundational elements of the architecture. The self-attention mechanism focuses on important dependencies between global particles, enabling the Global Feature Extractor to play a key role in enhancing performance. Notably, global interactions cannot exist independently of local features, which is locality principles underpin all particle-based methods. This further validates our Fluid Attention Block's pivotal value in enabling local-global synergy. Type-aware Embedding specifically strengthens fluid-solid coupling learning, establishing robust modeling foundation across the entire network architecture. Integration of relative position information via 3D-RoPE enhances spatial relationship details.

\setlength{\tabcolsep}{2mm}
\begin{table}[t]
  \centering
  \begin{tabular}{lcc}
    \toprule[0.6pt]
    \multirow{2}{*}{Method} & $n$-SE & MDE \\
    & (\(mm\)) & (\(g/cm^{3}\)) \\
    \midrule \midrule
    w/o Global Feature Extractor & 41.024 &  0.019\\
    w/o Local Feature Extractor & 75.073 & 0.057 \\
    w/o Type-aware Embedding & 84.462 &  0.066\\
    w/o CConv & 177.151 & 0.141 \\
    w/o ASCC & 93.524 &  0.073 \\
    w/o 3D-RoPE & 29.131 & 0.017  \\
    Ours & \textbf{24.442} & \textbf{0.008} \\
    \bottomrule[0.6pt]
  \end{tabular}
  \caption{Ablation study to quantify the contributions of key components in FluidFormer.}
  \label{tab:Ablation_Study}
\end{table}

\section{Conclusion}
We present FluidFormer, the first Transformer specifically designed for continuous fluid simulation. By innovatively combining convolution-based local feature extraction with attention-driven global context modeling, FluidFormer establishes a new paradigm for state-of-the-art fluid dynamics prediction. There are promising directions for future work, such as broader fluid simulation applications and framework extensions to rigid and deformable solids. We will release the code to facilitate such development.
\bibliography{my2026}
\end{sloppypar}
\end{document}